\documentclass[a4paper,11pt]{article}
\usepackage{pos}

\usepackage{braket} % bras and kets

\usepackage{float}

\usepackage{amsmath} % math

\usepackage{subfigure}
\usepackage{float}

\newcommand{\projectdir}{./}

\newcommand{\Njkf}{15}
\newcommand{\plotsdir}{\projectdir plots/}
\newcommand{\smartprod}[2]{
 \ifx#1#2
  #1^2
 \else
  #1 #2
 \fi
}

\newcommand{\FeynFld}{./jaxodraw/}
\newcommand{\FeynFldProp}{\FeynFld propagator_} % 2-point functions and corrections
\newcommand{\FeynFldBar}{\FeynFld baryons_} % baryon diagrams

\newcommand{\imeq}[2]{\vcenter{\hbox{\includegraphics[width=#1]{#2}}}}

\newcommand{\imeqDFT}[1]{\imeq{0.03\textwidth}{\FeynFld #1}} 

\newcommand{\imeqProp}[2]{\imeq{#1}{\FeynFldProp #2}}

\newcommand{\imeqBar}[2]{\imeq{#1}{\FeynFldBar #2}}
\newcommand{\imeqBarDFT}[1]{\imeqBar{0.13\textwidth}{#1}}

\usepackage{makecell}
\usepackage{float}

\newcommand{\EMsup}{\text{EM}}

\newcommand{\Csup}{\text{C}}
\newcommand{\PSEsup}{\Csup}

\newcommand{\sea}[1]{\textcolor{blue}{#1}}

\newcommand{\EMCsup}{\text{EMC}}

\newcommand{\MASSsup}{\text{M}}

\title{The neutron-proton mass difference}
\ShortTitle{$M_{n}-M_{p}$ in Lattice QCD+QED at LO in IB}

\author*[a,b]{Simone Romiti}
\affiliation[a]{Dipartimento di Matematica e Fisica, Università degli Studi Roma Tre, \\
Rome, Italy}
\affiliation[b]{Istituto Nazionale di Fisica Nucleare, Sezione di Roma Tre, Rome, Italy}
\emailAdd{simone.romiti@uniroma3.it}
\abstract{  
  We present a lattice
  calculation of the mass difference between neutron and proton,
  for which we find
  \mbox{
  $ M_n - M_p
   = \input{\projectdir QCDQED/Data/TeX/results/dM_np_feyn.dat}
   \, \text{MeV}
  $
  }.
  This is obtained at 1st order in the $QED$ coupling $\alpha_{EM}$ and in the mass difference between $u$ and $d$ quarks $\frac{m_d-m_u}{\Lambda_{QCD}}$. 
  We adopt a purely hadronic scheme to renormalize the theory and provide a prescription to separate the $QED$ and strong $IB$ contributions.
  The simulation is carried out using the ETMC gauge configurations with $N_f=2+1+1$ dynamical quarks. We extrapolate among $3$ values of the lattice spacing and pion masses in the range
  \mbox{$M_\pi \simeq 200 - 450 $ MeV}. 
}

\FullConference{%
 The 38th International Symposium on Lattice Field Theory, LATTICE2021
  26th-30th July, 2021
  Zoom/Gather@Massachusetts Institute of Technology
}

\begin{document}
\maketitle

\section{Introduction}
% introduction.tex

Today many lattice QCD calculations have reached the $O(1\%)$ precision level \cite{colangelo2011review},
requiring to include the Isospin Breaking (IB) Effects.
The leading terms of the latter come from the $\sim O(1\%)$ corrections of
$O(\hat{\alpha}_{EM})$ (\textit{QED}) 
and 
$O(\frac{\hat{m}_d - \hat{m}_u}{\Lambda_{QCD}})$
(\textit{strong IB}  or \textit{QCD}).
These can be taken into account simulating $u$ and $d$ quarks with different masses
and including QED in the action \cite{duncan1996electromagnetic, borsanyi2015ab}.
However, 
a convenient approach consists in expanding the path integral with respect to the isospin symmetry breaking parameters.
In this setup,
any observable in the full theory (QCD+QED) is the sum of its isosymmetric part and the IBEs.
This philosophy is the heart of the RM123 method \cite{de2012isospin, tantalo2013isospin, PhysRevD.87.114505} used in the present work.
For a given observable,
the slopes with respect to the IB couplings are found using the isosymmetric gauge configurations,
and the IB correction at Leading Order (LO) is a linear combination of these slopes with the appropriate charge factors and counterterms.
In this work the effect of IB on the hadronic spectum is investigated
with focus on $M_n-M_p$,
the mass difference between neutron and proton.
We also define a scheme for the separation of strong IB and QED contributions.
Though conventional,
this separation provides physical intuition about the sizes of the two effects.

% The calculation was done on the lattice 
% at LO in $\hat{\alpha}_{EM}$ 
% and $\frac{\hat{m}_d-\hat{m}_u}{\Lambda_{QCD}}$
% using the RM123 method.

We use a mixed action approach in the tmQCD regularization over the $N_f=2+1+1$ ETMC gauge configurations
\cite{carrasco2014up}.
In QCD+QED we adopt the electroquenched approximation and neglect QCD disconnected diagrams.
A purely hadronic renormalization scheme is adopted, 
using the masses of $\pi$s, $K$s and $\Omega^{-}$ as normalization parameters to set the scale, tune the counterterms and reach the physical point.

% outline dei risultati

The results found in this work are the following.
We obtain 
\mbox{
$ M_n - M_p
 = \input{\projectdir QCDQED/Data/TeX/results/dM_np_feyn.dat}
 \, \text{MeV}
$
,
}
\mbox{
$ 
(M_n - M_p)^{(QED)}
= \input{\projectdir QCDQED/Data/TeX/results/dM_np_feyn_QED.dat}
\, \text{MeV}                              
$
}
and
\mbox{
$
(M_n - M_p)^{(QCD)}
= \input{\projectdir QCDQED/Data/TeX/results/dM_np_feyn_QCD.dat}
\, \text{MeV}                              
$
}
.
We also get a prediction for the masses of nucleons,
\mbox{
$ M_{n}
 = \input{\projectdir QCDQED/Data/TeX/results/M_n.dat}
 \, \text{GeV}
$
}
,
\mbox{
$ M_{p}
 = \input{\projectdir QCDQED/Data/TeX/results/M_p.dat}
 \, \text{GeV}
$
}
and for the $\pi N$ sigma term
\mbox{
$
 \sigma_{\pi N}
 = \input{\projectdir isoQCD/Data/TeX/results/piN_sigma_term_MeV.dat}
 \, \text{MeV}                              
$
}
. 
% Finally, in the pion sector we get
% \mbox{
% $ M_{\pi^{+}}^2 - M_{\pi^{-}}^2
%  =
%  \input{\projectdir QCDQED/Data/TeX/results/dM_pi2.dat}
%  \, \text{MeV}^2
% $
% }
% .
The uncertainties are only statistical and obtained using the jackknife resampling technique.

The paper is organized as follows.
In sec. \eqref{sec:QCDQED.LO} we briefly recap the main features of the RM123 method and set our notation for the IBEs.
In sec. \eqref{sec:counterterms.separation} we discuss the tuning of counterterms and give our prescription for the separation between strong IB and QED.
In sec. \eqref{sec:details.simulation.analysis} and \eqref{sec:details.analysis} we respectively give the details of the simulations
and of the analysis for the above observables.
Finally in sec. \eqref{sec:summary.outlook} we give our summary and outlooks.

\section{QCD+QED at LO} 
\label{sec:QCDQED.LO}
% QCDQED_LO.tex
\newcommand{\introQCDQEDfeyn}[1]{\imeqProp{0.2\textwidth}{#1}}

In the continuum, 
the QCD+QED Lagrangian $\mathcal{L}$ can be written as:
\begin{align}
 \mathcal{L} & =                                        
 % \mathcal{L}_K
 % - m_u \bar{u} u
 % - m_d \bar{d} d
 % - e A_\mu
 % \left(
 % e_u \bar{u} \gamma_\mu u +
 % e_d \bar{d} \gamma_\mu d
 % \right)
 % \\
 %             & = \label{eq:Deltas.IB.Lagrangian.step.2} 
 % \mathcal{L}_K -
 % m_{ud} \, \bar{q} q
 % - \Delta m_{ud} \, \bar{q} \tau_3 q
 % + e A_{\mu} \bar{q} \gamma_\mu \left( \frac{\tau_3}{2} + \frac{1}{6} \right)  q
 % \\
 %             & = \label{eq:Deltas.IB.Lagrangian.step.3} 
 \mathcal{L}_0
 - \Delta m_{ud} \, \bar{q} \tau_3 q
 + e A_{\mu}
 \bar{q} \gamma_\mu \mathcal{Q}  q
 % \\
             % & = \label{eq:Deltas.IB.Lagrangian.step.4} 
 % =
 % \mathcal{L}_0
 % - \Delta m_{ud} \, \bar{q} \tau_3 q
 % + e A_{\mu}
 % J_\mu^{\mathcal{Q}}
 \quad .
\end{align}
The first term, $\mathcal{L}_0$, is symmetric under
$SU(2)_I$ (isoQCD theory) and chargeless under $U(1)_{EM}$,
while the rest are isospin breaking terms.
We write
\(m_u = m_{ud} - \Delta m_{ud}\),
\(m_d = m_{ud} + \Delta m_{ud}\).
and denote the up-down doublet with
\(q = (u,d)^{T}\). 
\(\tau_3\) is the third Pauli matrix and
$\mathcal{Q} = \frac{\tau_3}{2} + \frac{1}{6}$ .
At LO in IB we expand the path integral in 
$e^2$ and $\Delta m_{ud}$ 
(including $O(e^2)$ counterterms for the divergences of QED diagrams \cite{peskin1995introduction}).
The fine structure constant $\hat{\alpha}_{EM}$ renormalizes at higher orders with respect to our expansion, 
and we can safely use the value 
\mbox{
$
\alpha_{EM}
=e^2/(4\pi) 
=  1/137.035 999 084
$
} 
from \cite{10.1093/ptep/ptaa104}.
\\
On the lattice we use a mixed action approach as in \cite{PhysRevD.87.114505}.
This leads to the presence of counterterms for both the physical and critical masses, 
whose tuning is discussed in sec. \eqref{sec:counterterms.separation}.
The LIBE in a generic hadronic mass can then be written as:
\begin{equation}
  \Delta M_H = 
  \left[
  e^2 \bar{\Delta}^{\text{\EMsup}}
  \, + \sum_{f} a \Delta m_{f}^{cr} \bar{\Delta}_f^{\Csup} 
  \, + \sum_{f} a \Delta m_{f} \bar{\Delta}_f^{\MASSsup}
  \right]
  M_H
  \quad ,
\end{equation}
where the 
$\bar{\Delta}^{\EMsup}$
and 
$\bar{\Delta}_f^{x}$ 
\mbox{($x=\Csup, \MASSsup$)} (for a flavor $f$)
represent the slopes induced by the coupling in front of them.
Note that at the order we are working, 
these can be evaluated in the isosymmetric theory.
% ,
% i.e. over the same gauge configurations.
The slopes can be found individually from the corresponding corrections $\bar{\Delta}^{x}$ in those euclidean correlators whose isoQCD ground state has mass $M_H^{(0)}$.
In fact for large times 
\mbox{$C_H(t) \sim e^{-M_H t}$},
and it's easy to show that the mass slope's effective curve is given by:
\begin{equation}
  \label{eq:dM.from.dC.C0.no.back}
  \bar{\Delta}^{x} M_H
  = - \partial_t
  \left[ 
  \frac{\bar{\Delta}^{x} C_H(t)}{C_H^{(0)}(t)} 
  \right]
  \quad ,
\end{equation}
where $\partial_t f(t) = f(t+1)-f(t)$ (in lattice units).
At finite time extent this formula is valid in absence of backward signals,
i.e. for baryonic correlators with definite parity \cite{sasaki2000n, Lee_1999}.
In mesonic correlators however each forward signal is paired with a backward one,
and the above formula gets slightly modified \cite{PhysRevD.87.114505}.
%
% Here we use:
% %
% \begin{equation}
%   \label{eq:dM.from.dC.C0.yes.back}
%  {\bar{\Delta}^{x} M}_{\text{eff}} (t) =
%  \frac
%  { [\coth{(M_H^{(0)} \, (T/2 - t))}]^{(-)^p} }
%  {(T/2 - t) }
%  \left(
%  \frac{\bar{\Delta}^{x} C_H(t)}{C_H^{(0)}(t)} - \frac{\bar{\Delta}^{x} C_H(T/2)}{C_H^{(0)}(T/2)}
%  \right)
%  \quad ,
% \end{equation}
% %
% where $(-)^p$ is the sign of the correlator under time reversal.
% $M_H^{(0)}$ is the ground state mass extracted $C_H^{(0)}(t)$ at large times.
% In the following we'll use the notation 
% $-\partial_t$ as in eq. \eqref{eq:dM.from.dC.C0.no.back} 
% as shorthand also for mesons,
% implying that the appropriate formula is in use.
%
In this work we extract the mass slopes from a fit to a constant of these effective curves in their plateau intervals.
% \footnote{
% These formulas are valid in the leading exponential approximation for $C_H^{(0)}(t)$, 
% so the plateau of the slopes starts after the beginning of the effective curve $M_{(0)}_{\text{eff}}(t)$.
% }.
% In the following we'll use the notation 
% $-\partial_t$ as a shorthand also for mesons,
% implying that the appropriate formula is in use.
In \eqref{sec:LIBEs.spectrum.Feynman} 
are shown the expressions of the mass corrections in terms of Feynman diagrams.

We conclude this section saying that QED is introduced in a non-compact way \cite{duncan1996electromagnetic},
i.e. generating the photon field $A_{\mu}$ instead on the gauge link variables $E_\mu^{(f)}=\text{exp}(i q_f A_\mu)$),
and regularizing the infrared divergence in the photon propagator using the QED$_{\text{L}}$ prescription \cite{giusti2017leading}, 
i.e. removing the $\vec{k}=0$ mode.
QED on a torus $T \times L^3$ introduces Finite Volume Effects (FVEs) in the hadronic spectrum,
with polynomial behavior in $1/L$
\cite{10.1143/PTP.120.413, davoudi2014finite, FODOR2016245, borsanyi2015ab},
and the QED mass correction of an hadron obeys the following asymptotic formula:
\begin{equation}
  \label{eq:dM.QED.FVE}
  \Delta M(T,L) 
  \, \xrightarrow{T,L \to \infty} \, 
  \Delta M(\infty)
   - Q^2 \alpha_{EM}
    \left[
      \frac{\kappa}{2 M L}
      \left(
        1 + \frac{2}{ML}
      \right)
    \right]
  + O\left(\frac{\alpha_{EM}}{L^3}\right)
  \quad .
\end{equation}
$\kappa \approx 2.837297$, 
$Q$ is the charge of the hadron in units of $e$
and the $M$ in the denominators can be set equal to $M(\infty)$ at this order in $\alpha_{EM}$.
The terms $\sim 1/L$ and $1/L^2$ are universal, namely depend only on the electric charge and mass of the hadron,
with spin and structure-dependent terms starting only at higher orders.
% In the above expression we neglect terms which are exponentially suppressed or that fall faster than any power of $(MT)^{-1}$
% \footnote{
% This is justified because for our ensembles $T=2L$, 
% so that Finite Temperature Effects are subdominant with respect to FVEs.
% }
% .
% \\
At fixed ensemble,
we use this property to remove these universal FVEs from the mass corrections caused by the combination of diagrams coming from the interaction with the electromagnetic field (EM).

  \subsection{LIBEs in the hadronic spectrum}
  \label{sec:LIBEs.spectrum.Feynman}
  % LIBEs_Feynman.tex

%
The LIBEs in the hadronic spectrum can be drawn from Feynman diagrams.
For mesons the explicit expressions are provided in \cite{PhysRevD.87.114505}.
For baryons the drawing convention is analogous to \cite{de2012isospin}:
\begin{align}
  \bar{\Delta}^{\MASSsup} C_N^{(1)} &=
  -\imeqBarDFT{light3_MASS_q1_dir.pdf}
  +\imeqBarDFT{light3_MASS_q1_exch.pdf}
  \qquad , \quad ... \quad . \quad  ,
  \\
  \bar{\Delta}^{\PSEsup} C_N^{(1)} &=
  -\imeqBarDFT{light3_PSE_q1_dir.pdf}
  +\imeqBarDFT{light3_PSE_q1_exch.pdf}
  \qquad , \quad ... \quad . \quad  ,
  \\
  \bar{\Delta}^{\text{self}} C_N^{(1)} &=
  -\imeqBarDFT{light3_QED_self_q1_dir.pdf}
  -\imeqBarDFT{light3_QED_tad_q1_dir.pdf}
  +\imeqBarDFT{light3_QED_self_q1_exch.pdf}
  +\imeqBarDFT{light3_QED_tad_q1_exch.pdf}
  \qquad , \quad ... \quad . \quad  ,
  \\
  \bar{\Delta}^{\text{exch}} C_N^{(1)} &=
  -\imeqBarDFT{light3_QED_q2q3_dir.pdf}
  +\imeqBarDFT{light3_QED_q2q3_exch.pdf}
  \qquad , \quad ... \quad . \quad  ,
  \\
  \bar{\Delta}^{\text{loop}} C_N^{(1 \sea{f})} &=
  -\imeqBarDFT{light3_QED_loop_q1_dir.pdf}
  +\imeqBarDFT{light3_QED_loop_q1_exch.pdf}
  \qquad , \quad ... \quad . \quad  \quad .
\end{align}
The dots ``$...$'' are a shorthand to denote the other diagrams trivially obtained considering the insertions on the other legs.
The index $i=1,2,3$ in 
$\bar{\Delta}^{x} C_N^{(i)}$
corresponds to the quark propagator with the insertion of the current generating the slope 
(except for 
$\bar{\Delta}^{\text{exch}} C_N^{(i)}$, 
where $i$ denotes the leg unaffected by the photon exchange).
The $\imeqDFT{MASS_ins.pdf}$ and $\imeqDFT{PSE_ins.pdf}$ are the insertions of the scalar and pseudoscalar currents respectively.
For the $\Omega^{-}$ it's akin, 
differing for flavor content, spin $3/2$ projection,
and with a factor $2$ in front of the diagrams with crossed quark legs.
From the above equations we define the ratios
\mbox{
${\mathcal{R}}_i^{x} = -\partial_t [{\bar{\Delta}^{x} C_{\mathcal{R}}^{(i)}}/{C_{\mathcal{R}}^{(0)}}]$
}
for $x \in \{\MASSsup, \PSEsup, \text{self}, \text{exch}\}$
and 
\mbox{
${\mathcal{R}}_{i \sea{f} }^{\text{loop}} = -\partial_t [{\bar{\Delta}^{x} C_{\mathcal{R}}^{(i \sea{f})}}/{C_{\mathcal{R}}^{(0)}}]$
},
where
${\mathcal{R}} = N, \Omega$
.
The LIBEs in the nucleon doublet then assume the following form 
(see eq. \eqref{eq:dM.from.dC.C0.no.back}):
\begin{equation}
  \label{eq:dMn.Feynman}
  \begin{aligned}
  \Delta M_n
  =& 
  - \Delta m_u {N}_1^{\MASSsup}
  - \Delta m_d {N}_2^{\MASSsup}
  - \Delta m_d {N}_3^{\MASSsup}
  + \Delta m_u^{(cr)} {N}_1^{\PSEsup}
  + \Delta m_d^{(cr)} {N}_2^{\PSEsup}
  + \Delta m_d^{(cr)} {N}_3^{\PSEsup}
  \\
  &
  + q_u^{2} {N}_1^{\text{self}}
  + q_d^{2} {N}_2^{\text{self}}
  + q_d^{2} {N}_3^{\text{self}}
  + q_u q_d {N}_{3}^{\text{exch}}
  + q_u q_d {N}_{2}^{\text{exch}}
  + q_d^{2} {N}_{1}^{\text{exch}}
  \\
  &
  + 
  \sum_{\sea{f \in (sea)}}
  \sea{q_f}
  \left[
  q_u {N}_{1 \sea{f} }^{\text{loop}}
  + q_d {N}_{2 \sea{f} }^{\text{loop}}
  + q_d {N}_{3 \sea{f} }^{\text{loop}}
  \right]
  + [\text{isosymm. vac. pol. diag.}]
  \quad ,
\end{aligned}
\end{equation}
and
\begin{equation}
  \begin{aligned}
    \label{eq:dMp.Feynman}
  \Delta M_p
  =& 
  - \Delta m_d {N}_1^{\MASSsup}
  - \Delta m_u {N}_2^{\MASSsup}
  - \Delta m_u {N}_3^{\MASSsup}
  + \Delta m_d^{(cr)} {N}_1^{\PSEsup}
  + \Delta m_u^{(cr)} {N}_2^{\PSEsup}
  + \Delta m_u^{(cr)} {N}_3^{\PSEsup}
  \\
  &
  + q_d^{2} {N}_1^{\text{self}}
  + q_u^{2} {N}_2^{\text{self}}
  + q_u^{2} {N}_3^{\text{self}}
  + q_d q_u {N}_{3}^{\text{exch}}
  + q_d q_u {N}_{2}^{\text{exch}}
  + q_u^{2} {N}_{1}^{\text{exch}}
  \\
  &
  + 
  \sum_{\sea{f \in (sea)}}
  \sea{q_f}
  \left[
  q_d {N}_{1 \sea{f} }^{\text{loop}}
  + q_u {N}_{2 \sea{f} }^{\text{loop}}
  + q_u {N}_{3 \sea{f} }^{\text{loop}}
  \right]
  + [\text{isosymm. vac. pol. diag.}]
  \quad .
\end{aligned}
\end{equation}
The neutron-proton mass difference at LO is:
\begin{equation} \label{eq:dMnp.Feynman}
  \begin{aligned}
  M_n-M_p
  =
  & 
  - 2 \Delta m_{ud} 
  [
  {N}_1^{\MASSsup}
  - {N}_2^{\MASSsup}
  - {N}_3^{\MASSsup}
  ]
  + 2 \Delta m_{ud}^{(cr)} 
  [
  {N}_1^{\PSEsup}
  - {N}_2^{\PSEsup}
  - {N}_3^{\PSEsup}
  ]
  \\
  &
  + (q_u^{2}-q_d^{2}) [{N}_1^{\text{self}}
  - {N}_2^{\text{self}}
  - {N}_3^{\text{self}}
  - {N}_{1}^{\text{exch}}
  ]
  \\
  &
  % \\
  % &
  + 
  \sum_{\sea{f \in (sea)}}
  \sea{q_f}
  (q_u-q_d)\left[
  {N}_{1 \sea{f} }^{\text{loop}}
  - {N}_{2 \sea{f} }^{\text{loop}}
  - {N}_{3 \sea{f} }^{\text{loop}}
  \right]
  \quad .
\end{aligned}
\end{equation}
Finally the IB correction to $M_{\Omega^{-}}$ is:
\begin{equation}
  \begin{aligned}
  \Delta M_{\Omega^{-}}
  =& 
  - \Delta m_s 
  [
  {{\Omega}}_1^{\MASSsup}
  + {{\Omega}}_2^{\MASSsup}
  + {{\Omega}}_3^{\MASSsup}
  ]
  + \Delta m_s^{(cr)} 
  [
  {{\Omega}}_1^{\PSEsup}
  +{{\Omega}}_2^{\PSEsup}
  +{{\Omega}}_3^{\PSEsup}
  ]
  % \\
  % &
  \\
  &
  + q_s^2 
  [
  {{\Omega}}_1^{\text{self}}
  + {{\Omega}}_2^{\text{self}}
  + {{\Omega}}_3^{\text{self}}
  + {{\Omega}}_{3}^{\text{exch}}
  + {{\Omega}}_{2}^{\text{exch}}
  + {{\Omega}}_{1}^{\text{exch}}
  ]
  \\
  &
  + 
  \sum_{\sea{f \in (sea)}}
  \sea{q_f} q_s
  \left[
  {{\Omega}}_{1 \sea{f} }^{\text{loop}}
  + {{\Omega}}_{2 \sea{f} }^{\text{loop}}
  + {{\Omega}}_{3 \sea{f} }^{\text{loop}}
  \right]
  + [\text{isosymm. vac. pol. diag.}]
  \quad .
\end{aligned}
\end{equation}

In this work we neglect the (disconnected) isosymmetric vacuum polarization diagrams,
and work in the electroquenched approximation: 
sea quarks are neutral with respect to the photon field, i.e. all diagrams with photons attached to quark loops vanish. 
This gives the numerical advantage of not having to evaluate the LIBEs also in the sea quark determinant.
% , 
% and recycle the old isosymmetric gauge configurations. 

%

\section{Counterterms and separation between QCD and QED}
\label{sec:counterterms.separation}
% tuning.tex 

In tmQCD the inclusion of QED introduces counterterms to the critical and physical masses.\\
In this work the former are tuned from the infinite volume limit of the PCAC Ward Identity (WI) integrated over the $3$ spatial directions.
Our values of $a m_0^{cr}$ have been already tuned to get maximal twist in absence of IB \cite{baron2010light}.
Here we require to preserve maximal twist
(and hence the $O(a)$ improvement \cite{gattringer2009quantum}) 
also at $O(e^2)$.
Therefore, for each flavor, 
% at fixed ensemble 
we find the bare counterterms 
$a \Delta m_f$
from a fit to a constant of the following condition in its plateau region:
\begin{equation}
	0 
  = \Delta {m}^{PCAC}_f (t) =
	\Delta^{\EMCsup} \left( 
	\frac{\partial_{t} \braket{A^a_{4} (t) \, P^a(0)}}{\braket{P^a (t) \, P^a (0)}}
	\right)
	\quad ,
\end{equation}
where
% \footnote{
% Here 
% $\psi_f$
% is the spinor doublet for the flavor $f$ in the physical basis.
% }
\mbox{$P^a = \sum_{\vec{x}}{\bar{\chi}}_{f}(x) \gamma^5 \frac{\tau^a}{2} \chi_{f}(x)$}
and
\mbox{$A_\mu^a = \sum_{\vec{x}} {\bar{\chi}}_{f}(x) \gamma_\mu \gamma^5 \frac{\tau^a}{2} \chi_{f}(x)$}
in the twisted basis.
% The variation $\Delta$ is induced by photon interactions and the counterterms $\Delta m_f^{(cr)}$.
The tuning is done at fixed ensemble,
and we sum the $\EMsup$ effective correction with the effect of critical mass counterterms (C) getting the variation $\Delta M^{\EMCsup}$.

The physical mass counterterms are then tuned as follows.
We define the physical point of both isoQCD and QCD+QED from the ratios:
\begin{equation}
	\label{eq:rs.rl.rp.definition}
 r_s     =                                                
 \frac{
 2(M_{K^+}^2 + M_{K^0}^2)
 -(M_{\pi^+}^2+M_{\pi^0}^2)
 }{2 M_{\Omega^{-}}^2}    
 \quad , \quad 
 r_\ell  = \frac{M_{\pi^+}^2+M_{\pi^0}^2}{2 M_{\Omega^{-}}^2} 
 \quad , \quad 
 r_p     = \frac{M_{K^+}^2}{M_{\Omega^{-}}^2}                 
 \quad ,
\end{equation}
imposing 
$r_s=r_s^{(exp)}$,
$r_\ell=r_\ell^{(exp)}$ and
$r_p=r_p^{(exp)}$.
This also implies, by definition, that their total IB corrections vanishes.
Expanding according to the LO corrections to the masses (see sec. \eqref{sec:LIBEs.spectrum.Feynman}), we have:
\begin{equation}
 r_i    = r_i^{(0)} +    
 \sum_{f\in(u,d,s)} a \Delta m_f {\bar{\Delta}^{\MASSsup}_f r_i}+
 {\Delta} r_i^{\EMCsup}
 \quad \, (i = s, \ell, p) \quad , 
\end{equation}
%
% where $\bar{\Delta}^{\MASSsup} r_i ^{(f)}$ 
% ($i=s,\ell,p$) 
% is the slope generated by the counterterm 
% $a \Delta m_f$, 
% and $(EMC)$ denotes the combination of induced by photons and the critical mass counterterms (already tuned from the PCAC WI).
% %
so that at our physical point,
where $r_i=r_i^{(0)}$,
the solution to the above system defines the counterterms as:
\begin{gather}\label{eq:QCDQED.tuning.system.solution}
 \begin{bmatrix} a \Delta m_u \\a \Delta m_d \\ a \Delta m_s \end{bmatrix}
 = -
 \begin{bmatrix}
  \bar{\Delta}^{\MASSsup}_{u} r_s    & \bar{\Delta}^{\MASSsup}_{d} r_s    & \bar{\Delta}^{\MASSsup}_{s} r_s    \\
  \bar{\Delta}^{\MASSsup}_{u} r_\ell & \bar{\Delta}^{\MASSsup}_{d} r_\ell & \bar{\Delta}^{\MASSsup}_{s} r_\ell \\
  \bar{\Delta}^{\MASSsup}_{u} r_p    & \bar{\Delta}^{\MASSsup}_{d} r_p    & \bar{\Delta}^{\MASSsup}_{s} r_p    
 \end{bmatrix}^{-1}
 \begin{bmatrix}
  \Delta r_s ^{\EMCsup}      \\
  \Delta {r_\ell} ^{\EMCsup} \\
  \Delta r_p ^{\EMCsup}      \\
 \end{bmatrix}
 \quad .
\end{gather}
These equations give the counterterms at the isoQCD physical point of $a m_s$ and $a m_\ell$ (or equivalently of $r_s$ and $r_\ell$).
In the analysis however the $a \Delta m_f$ are found at fixed simulated light quark mass,
after the interpolation of the slopes among the $2$ values of $a m_s$ to the physical point of $r_s$.
These $a \Delta m_f$ are used to evaluate the other observables,
which are then extrapolated to $L \to \infty$, $a \to 0$ and $r_\ell = r_\ell^{(exp)}$ over all the ensembles.
For each observable $O$ this is just an extrapolation in separate steps, 
done on the slice $r_s=r_s^{(exp)}$ of the hyper-surface $O(r_s, r_\ell, L, a)$.

\begin{figure}
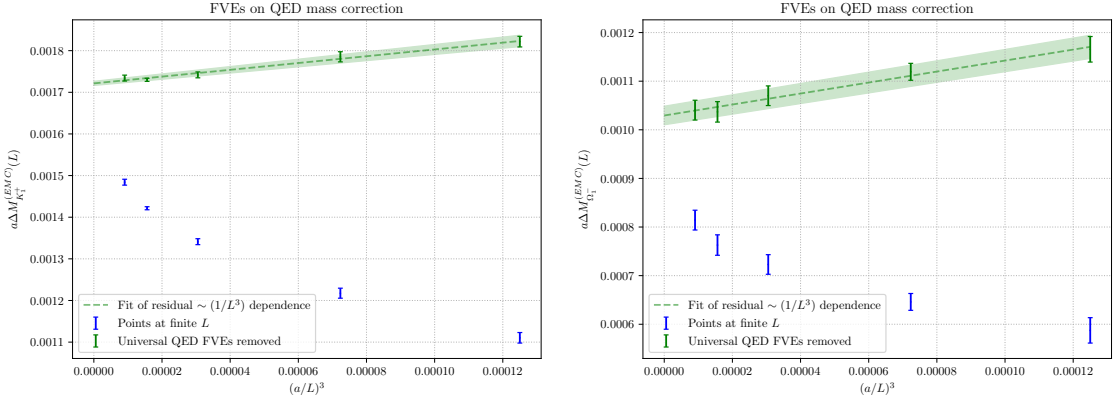

    \begin{center}
			\includegraphics[width=0.49\textwidth]{\plotsdir QCDQED/FVE/QED_L/dM_K_plus_EMC_s1.pdf}
			\includegraphics[width=0.49\textwidth]{\plotsdir QCDQED/FVE/QED_L/dM_Omega_EMC_s1.pdf}
    \end{center}
  \caption{
    \begin{small}
			$\EMCsup$ corrections for the $K^{+}$ and $\Omega^{-}$ obtained from the $1$st of the $2$ simulated values of $a m_s$,
			for the A40.XX ensembles. 
      The blue points are the $\EMCsup{}$ corrections in lattice units at finite volume. 
      % The blue band is the theoretical curve corresponding to the universal effects,
      % with the uncertainty coming from the mass of the corresponding hadron.
      The green points are the values after the correction of the universal QED FVEs.
      The green band is a fit of the latter with an ansatz of the form 
      \mbox{$A + B\left(\frac{a}{L}\right)^3$}, 
      with $A$ and $B$ free parameters of the fit.
    \end{small}
   }
  \label{fig:QCDQED.FVE.universal}
\end{figure}

We remark that,
after the subtraction of the universal $1/L$ and $1/L^2$ corrections of eq. \eqref{eq:dM.QED.FVE},
these $\Delta m_f$ contain residual $O(\alpha_{EM}/L^3)$ FVEs from QED.
% but not the universal terms of eq. \eqref{eq:dM.QED.FVE}.
These have been removed before the tuning from the $\EMCsup{}$ corrections.
Fig. \eqref{fig:QCDQED.FVE.universal} shows the behavior of 2 of our $\EMCsup{}$ corrections using the A40.XX ensembles,
which differ only for the volume.

We can now provide a prescription for the separation of strong IB (QCD) and QED.
Here we implement the separation as in \cite{PhysRevD.87.114505}, namely we write:
\begin{equation}
\Delta m_{ud} =
\frac{m_{d}-m_{u}}{2} =
\Delta m_{ud}^{(QCD)} + \Delta m_{ud}^{(QED)}
 =  
 \frac{\Delta \hat{m}_{ud}}{Z_P^{(0)}}
 + \frac{(q_d^2-q_u^2)}{32 \pi^2}
  \left[6 \log{(a \mu)} - 22.595 \right] m_\ell^{(0)}  
	\quad .
\end{equation}
With the notation of sec. \eqref{sec:LIBEs.spectrum.Feynman},
the contributions to the mass difference $M_n-M_p$ then read:
\begin{equation}
	\label{eq:dMnp.QCD.expression}
  \begin{aligned}
  (M_n-M_p)^{(QCD)}
  = 
  - 2 \Delta m_{ud}^{(QCD)} 
  [
  {\mathcal{N}}_1^{\MASSsup}
  - {\mathcal{N}}_2^{\MASSsup}
  - {\mathcal{N}}_3^{\MASSsup}
  ]
	\quad ,
\end{aligned}
\end{equation}
\begin{equation}
	\label{eq:dMnp.QED.expression}
  \begin{aligned}
  (M_n-M_p)^{(QED)}
  & = 
  - 2 \Delta m_{ud}^{(QED)} 
  [
  {\mathcal{N}}_1^{\MASSsup}
  - {\mathcal{N}}_2^{\MASSsup}
  - {\mathcal{N}}_3^{\MASSsup}
  ]
  + 2 \Delta m_{ud}^{(cr)} 
  [
  {\mathcal{N}}_1^{\PSEsup}
  - {\mathcal{N}}_2^{\PSEsup}
  - {\mathcal{N}}_3^{\PSEsup}
  ]
  \\
  &
  + (q_u^{2}-q_d^{2}) [{\mathcal{N}}_1^{\text{self}}
  - {\mathcal{N}}_2^{\text{self}}
  - {\mathcal{N}}_3^{\text{self}}
  - {\mathcal{N}}_{23}^{\text{exch}}
  ]
  \\
  &
  + 
  \sum_{\sea{f \in (sea)}}
  \sea{q_f}
  (q_u-q_d)\left[
  {\mathcal{N}}_{1 \sea{f} }^{\text{loop}}
  - {\mathcal{N}}_{2 \sea{f} }^{\text{loop}}
  - {\mathcal{N}}_{3 \sea{f} }^{\text{loop}}
  \right]
  \quad .
\end{aligned}
\end{equation}
The physical interpretation is that the neutron tends to
be heavier because \(m_d > m_u\). 
The \(u\) however has a bigger electric charge, giving an higher electromagnetic self
energy to the proton.
In nature it happens that these
effects are of the same order of magnitude,
canceling
almost exactly and leaving a small mass difference, $O(1 \, \text{MeV})$, compared to their masses,
$O(1 \, \text{GeV})$.
%
% Therefore in our analysis
% $(M_n-M_p)^{(QED)}$ is obtained replacing 
% \(\Delta m_{ud}\) with \(\Delta m_{ud}^{(QED)}\) in the LO expression.
% The strong IB effect is found considering only the $\bar{\Delta}^{\MASSsup}$ slope multiplied by \(\Delta m_{ud}^{(QCD)}\).
%
%
We conclude remarking, however, that the above separation is a matter of prescription, 
since it's arbitrary to choose which finite terms go in the divergent term $\Delta m_{ud}^{(QED)}$.

\section{Details of simulation and analysis}
\label{sec:details.simulation.analysis}
Our correlators are evaluated over the 
$N_f=2+1+1$ ETMC gauge configurations \cite{carrasco2014up}
with twisted mass quarks at maximal twist.
We adopt a unitary setup in the light sector,
and a mixed action approach for the strange and charm quarks
which in the valence are regularized as Osterwalder-Seyler fermions \cite{OSTERWALDER1978440}.
In tab. \eqref{tab:masses&simudetails} are reported the ensembles details. 
% In a first run (isoQCD) we considered the isosymmetric theory for $3$ values of the valence quark mass, 
% $a m_s$, 
% interpolating the observables to its physical point.
% In a second run (QCD+QED), 
% we evaluated also the IB slopes, 
% simulating $2$ values of $a m_s$ in order to propagate the uncertainty on the physical point found in isoQCD.
% In tab. \eqref{tab:ams.isoQCD.QCDQED} are given the values of the valence strange quark mass $a m_s$ for the two runs.
% %
% \begin{table}[p]
%   \begin{center}
%   \begin{tabular}{|c|c|c|}
%     \hline \\[-1em]
%     $\beta$ & $am_s$ (isoQCD) & $am_s$ (QCD+QED) \\
%     \hline \\[-1em]
%     $1.90$  & 
%     \makecell{$0.0220$, $ 0.0260$,\\$ 0.0300$} & \makecell{$0.0242$, $0.0261$} \\ 
%     \hline \\[-1em]
%     $1.95$  & 
%     \makecell{$0.0136$, $0.0161$, \\$0.0186$} & \makecell{$0.0216$, $0.0230$} \\ 
%     \hline \\[-1em]
%     $2.10$  & 
%     \makecell{$0.0220$, $ 0.0260$,\\$ 0.0300$} & \makecell{$0.0176$, $0.0186$} \\ 
%     \hline
%   \end{tabular}
%   \end{center}
%   \caption{
%   Values of $a m_s$ in runs $1$st (isoQCD) and $2$nd (QCD+QED) run.}
% \end{table}
%
%
\begin{table}
\begin{center}
\footnotesize
\renewcommand{\arraystretch}{1.20}
\begin{tabular}{||c|c|c|c|c|c|c|c|c||}
\hline
Ensemble & $\beta$ & $V / a^4$         & $a m_{sea}=a m_\ell$   & $a m_s$ & $a m_\sigma$ & $a m_\delta$ & $\kappa$    & $N_{cfg}$ \\
\hline \hline
$A30.32$  & $1.90$ & $32^{3}\times 64$ & $0.0030$ & \makecell{$0.0242$, $0.0261$}  & $0.15$  & $0.19$  & $0.163272$  & $150$ \\
$A40.32$  &        &                   & $0.0040$ &                                &         &         & $0.163270$  & $150$ \\
$A50.32$  &        &                   & $0.0050$ &                                &         &         & $0.163267$  & $150$ \\
\hline 
$A40.20$  & $1.90$ & $20^{3}\times 48$ & $0.0040$ & \makecell{$0.0242$, $0.0261$}  & $0.15$  & $0.19$  & $0.163270$  & $150$ \\
\hline 
$A40.24$  & $1.90$ & $24^{3}\times 48$ & $0.0040$ & \makecell{$0.0242$, $0.0261$}  & $0.15$  & $0.19$  & $0.163270$  & $150$ \\
$A60.24$  &        &                   & $0.0060$ &                                &         &         & $0.163265$  & $150$ \\
$A80.24$  &        &                   & $0.0080$ &                                &         &         & $0.163255$  & $150$ \\
$A100.24$ &        &                   & $0.0100$ &                                &         &         & $0.163260$  & $150$ \\
\hline 
$A40.48$  & $1.90$ & $48^{3}\times 96$ & $0.0040$ & \makecell{$0.0242$, $0.0261$}  & $0.15$  & $0.19$  & $0.163270$  & $90$  \\
\hline 
$A40.40$  & $1.90$ & $40^{3}\times 80$ & $0.0040$ & \makecell{$0.0242$, $0.0261$}  & $0.15$  & $0.19$  & $0.163270$  & $150$  \\
\hline \hline
$B25.32$  & $1.95$ & $32^{3}\times 64$ & $0.0025$ & \makecell{$0.0216$, $0.0230$}  & $0.135$ & $0.170$ & $0.1612420$ & $150$ \\
$B35.32$  &        &                   & $0.0035$ &                                &         &         & $0.1612400$ & $150$ \\
$B55.32$  &        &                   & $0.0055$ &                                &         &         & $0.1612360$ & $150$ \\
$B75.32$  &        &                   & $0.0075$ &                                &         &         & $0.1612320$ & $75$  \\
\hline
$B85.24$  & $1.95$ & $24^{3}\times 48$ & $0.0085$ & \makecell{$0.0216$, $0.0230$}  & $0.135$ & $0.170$ & $0.1612312$ & $150$ \\
\hline \hline
$D15.48$  & $2.10$ & $48^{3}\times 96$ & $0.0015$ & \makecell{$0.0176$,  $0.0186$} & $0.12$ & $0.1385$ & $0.156361$  & $90$  \\ 
$D20.48$  &        &                   & $0.0020$ &                                &        &          & $0.156357$  & $90$  \\
$D30.48$  &        &                   & $0.0030$ &                                &        &          & $0.156355$  & $90$  \\
 \hline   
\end{tabular}
\renewcommand{\arraystretch}{1.0}
\end{center}
\caption{
Parameters of the ensembles used in this work. The space-time volume is reported in the format $L^3 \times T$. The bare values for $\beta$, sea and valence quark masses and hopping parameter $\kappa$ are reported. 
$a m_\sigma$ and $a m_\delta$ are the parameters which determine the renormalized strange and charm sea quark masses according to eq. (9) of \cite{carrasco2014up}.
In the rightmost column there are the number of analyzed gauge configurations.}
\label{tab:masses&simudetails}
\end{table}
The statistical uncertainty on our observables was propagated using the jackknife re-sampling technique with $\Njkf$ jackknifes for each ensemble.
% Gaussian smearing
Gaussian smearing was applied to quark fields according to \cite{de2012isospin}, 
with the parameter $\alpha_g$ optimized as in \cite{alexandrou2008light}.
Some numerical testing lead us to the choice of $n_g = 50$ steps on the source of our correlators, as an appropriate middle ground for a soon plateau and moderate noise in the signal.
%stochastic sources
In order to reduce the noise in our correlator, we used $16$ stochastic sources \cite{foster1999quark} for the numerical inversion of the Dirac operator, compatibly with our computational resources.

The isosymmetric limit of hadronic masses $M_H$
are found from the large time behavior of $\vec{p}=\vec{0}$ correlators
% :
% %
% \begin{equation}
%   C_T(t) 
%   = \sum_{\vec{x}} C_T(t, \vec{x})
%   = \sum_{\vec{x}} \braket{\mathcal{O}(t,\vec{x}) \mathcal{O}^\dagger(0, \vec{0})}_T
%   \quad ,
% \end{equation}
% %
% where $\mathcal{O}$ is the interpolating operator 
with $M_H^{(0)}$ as the ground state.
These values are found from a fit to a constant of the effective mass curves \cite{gattringer2009quantum} in their plateau intervals \footnote{
As a consistency check, 
we also verified the results with the leading exponential fit and the ODE method \cite{PhysRevD.100.054515}.
},
and we do the same also for the effective slopes curves 
(see sec. \eqref{sec:QCDQED.LO}).

Our lattice ensembles are not at the physical point, 
requiring to extrapolate among the ensembles.
% so we to extrapolate among the values obtained for each of them.
This is done in an hadronic scheme,
in terms of the ratios defined in eq. \eqref{eq:rs.rl.rp.definition}. 
$M_{\Omega^{-}}$, $(M_{\pi^+}^2+M_{\pi^+}^2)$, $M_{K^+}^2$ and $M_{K^0}^2$ are used to tune the parameters $a$, $a m_u$, $a m_d$ and $a m_s$,
with the counterterms $a \Delta m_f$ defined at the isoQCD physical point (see sec. \eqref{sec:counterterms.separation}).
The lattice spacings $a_{\beta(i)} = 0.1011(10), 0.09029(77), 0.06834(63)$ fm 
at 
$\beta = 1.90, 1.95, 2.10$ 
respectively, 
% (see tab. \eqref{tab:QCDQED.ai.fit.Ll}) 
% %
% \begin{table}
%   \begin{center}
%     \begin{tabular}{c|c|c|c}
%       \input{\projectdir QCDQED/Data/fit_Lla/TeX/ai.dat}
%     \end{tabular}
%   \end{center}
%   \caption{
%   Lattice spacings obtained from $aM_{\Omega^{-}}$ 
%   with the simultaneous fit over $L$ and $r_\ell$.
%   }
%   \label{tab:QCDQED.ai.fit.Ll}
% \end{table}
% %
are found from the extrapolation of $a M_{\Omega^{-}}$ among the ensembles,
whose values are fitted with the following polynomial ansatz:
\begin{equation}
 \label{eq:aMOmega.ansatz.Ll.QCDQED}
 (aM_{\Omega^-})_i (L, r_\ell) \, = \,
 a_{\beta(i)} \, M_{\Omega^{-}}^{(exp)}
 \left[
  1 + \,
  c_{L} \frac{\alpha_{EM}}{L^3} + \,
  c_\ell \, \Delta r_\ell \, + \,
  c_{\ell}^{(2)} \, \Delta r_\ell^2
  \right]
 \quad .
\end{equation}
$\Delta r_\ell=r_\ell-r_\ell^{exp}$
and
$c_L$, $c_\ell$, $c_\ell^{(2)}$ and the $a_{\beta(i)}$ ($i=1,2,3$) are free parameters of the fit.
The numerically leading FVE is taken into account with the $\sim 1/L^3$ term,
coming from the residual QED FVEs in $a \Delta m_s$ and $\Delta M_{\Omega^{-}}^{\EMCsup{}}$.
% In fig. \eqref{fig:QCDQED.aMOmega.fit.Ll} is shown the plot of the simultaneous extrapolation.
% %
% \begin{figure}
%  \begin{center}
%   \includegraphics[width=0.7\textwidth]{\plotsdir QCDQED/fit_Lla/scale_setting_Ll/M_Omega.pdf}
%  \end{center}
%  \caption{Extrapolation of $aM_{\Omega^{-}}$ in QCD+QED over all the ensembles according to the ansatz given in eq. \eqref{eq:aMOmega.ansatz.Ll.QCDQED}.
%   In faint grey are shown the points at finite volume, 
%   and the colored dashed lines (with error bands) are the curves extrapolated to \mbox{$L \to \infty$}.
%  }
%  \label{fig:QCDQED.aMOmega.fit.Ll}
% \end{figure}
% %

Each observable is extrapolated among $2$ values of $a m_s$ and the $a m_\ell$ of tab. \eqref{tab:masses&simudetails} to the isoQCD physical point 
in terms of the ratios $r_s$ and $r_\ell$ of eq. \eqref{eq:rs.rl.rp.definition}.
FVEs are fitted using asymptotic formulas from ChPT (isoQCD FVEs) \cite{colangelo2004pion, colangelo2005finite,colangelo2010finite} and QED at finite volume \cite{
10.1143/PTP.120.413,davoudi2014finite, FODOR2016245,borsanyi2015ab},
% However, the former are exponentially suppressed while the latter have a power law behavior starting at $O(\alpha_{EM}/L^3)$ \footnote{
while discretization effects are included with $O(a^2)$ terms
in virtue of the $O(a)$ improvement provided by maximal twist.
% We don't need to fit the universal QED FVEs of $O(\alpha_{EM}/L)$ and $O(\alpha_{EM}/L^2)$,
% which are removed with eq. \eqref{eq:dM.QED.FVE}.
% }.
%
\section{Nucleons spectrum}
\label{sec:details.analysis}
The masses of nucleons are found for each ensemble, 
summing the isosymmetric part $M_N$ to the IB corrections given in eqs. \eqref{eq:dMn.Feynman} and \eqref{eq:dMp.Feynman}.
We fit $M_n$ and $M_p$
% their average 
% \mbox{$M_{np}=(M_n+M_p)/2$}
according to the following ansatz inspired by ChPT \cite{alexandrou2009low}:
\begin{equation}
  \begin{aligned}
    M_{n/p}(L, r_\ell, a) = 
    A_{n/p} \,
    & 
    \left[
    1
    + \alpha_{EM} \frac{c_3^{({n/p})}}{{L}^3}
    + c_a^{({n/p})}  a^2
    + c_\ell^{({n/p})}  r_\ell
    + c_{3/2}^{({n/p})} r_\ell^{3/2}
 \right]
 \quad ,
  \end{aligned}
\end{equation}
where the coefficients 
$A_{n/p}$, ... 
are free parameters of the fit.
Similarly,
The mass difference $M_n-M_p$ is found from eq. \eqref{eq:dMnp.Feynman}, 
and fitted with a simple polynomial ansatz:
\begin{equation}
  \label{eq:dMnp.ansatz}
  \begin{aligned}
 (M_n-M_p)(L, r_\ell, a) =
 A \,
 & 
 \left[
 1
 + \alpha_{EM} \frac{c_3}{{L}^3}
 + c_a  a^2
 + c_\ell  r_\ell
 \right]
  \quad ,
\end{aligned}
\end{equation}
where the coefficients $A$, ..., are free parameters of the fit.
We account for the volume dependence with an $O(\alpha_{EM}/L^3)$ term, 
arising from the structure-dependence and the residual dependence in physical mass counterterms.
In both cases higher orders of $1/L$ and the isoQCD FVEs are found to be numerically negligible at our level of precision.

\begin{figure}
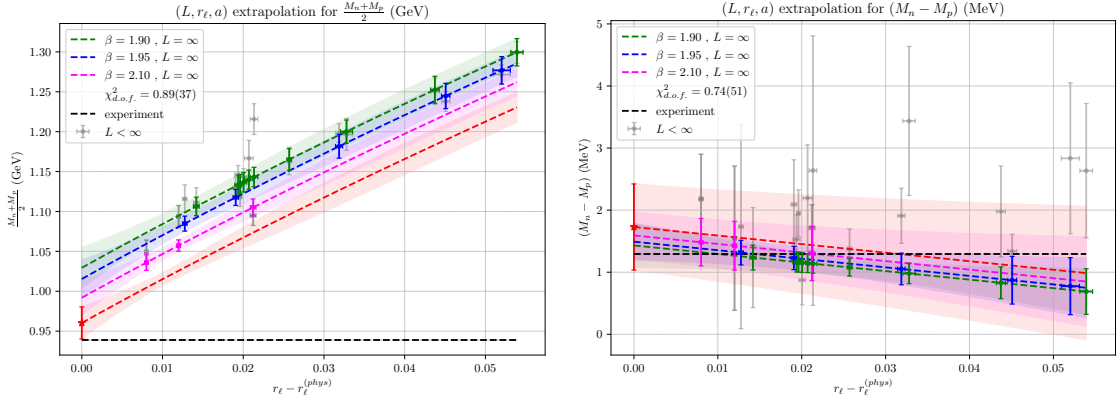

  \begin{center}
  \includegraphics[width=0.49\textwidth]{\plotsdir QCDQED/fit_Lla/M_np.pdf}
 \includegraphics[width=0.49\textwidth]{\plotsdir QCDQED/fit_Lla/dM_np_feyn.pdf}
  \end{center}
 \caption[]{
  Extrapolation of $(M_n+M_p)/2$ (left panel) and $M_n-M_p$ (right panel).
  Grey points are the values at finite volume.
  The colored points and lines correspond to the limit $L\to\infty$.
  The red curves are the continuum and infinite volume limits.
  The final predictions at $r_\ell=r_\ell^{(\text{phys})}$ are marked on the left.
  The horizontal black lines are the experimental values.
  % For the coefficients we find:
  % $A_{np}=1.82(67)$ MeV, 
  % $c_\ell=-7.7(7.9)$.
  % $c_a=-6.5(12.1)$, 
  % $\alpha_{EM} c_3=3.03(3.08)$, 
 }
 \label{fig:Mnp.and.dMnp.extrapolation.QCDQED}
\end{figure}

In fig.
\eqref{fig:Mnp.and.dMnp.extrapolation.QCDQED},
% \eqref{fig:dMnp.strongIB.extrapolation.QCDQED}
% and
% \eqref{fig:dMnp.QED.extrapolation.QCDQED}
we show the plot
of the extrapolation
for the average mass 
\mbox{$M_{np}=(M_n+M_p)/2$} 
and 
\mbox{$M_n-M_p$}.
% for $M_n-M_p$ and its $2$ contributions coming from strong IB and QED.
%
%
%
% %
% In fig. \eqref{fig:Mn.Mp.extrapolation.QCDQED}
% we show the plots of the extrapolation for $M_n$ and $M_p$ respectively.
% %
% \begin{figure}
%   \begin{center}
%   \includegraphics[width=0.49\textwidth]{\plotsdir QCDQED/fit_Lla/M_n.pdf}
%     \includegraphics[width=0.49\textwidth]{\plotsdir QCDQED/fit_Lla/M_p.pdf}
%   \end{center}
%  \caption[]{
%   Extrapolation of $M_n$ and $M_p$ in QCD+QED at LO.
%   Grey points are the values at finite volume.
%   The colored points and lines correspond to the limit $L\to\infty$.
%   The red curve is the continuum and infinite volume limit.
%   The final prediction at $r_\ell=r_\ell^{(\text{exp})}$ is marked on the left.
%   The horizontal black line is the experimental value.
%  }
%  \label{fig:Mn.Mp.extrapolation.QCDQED}
% \end{figure}
%
Our predictions are the following, 
whith the uncertainties being statistical.
We find \cite{Mohr:2015ccw}:
% %
% \begin{equation}
%   M_{np} = \frac{M_n+M_p}{2} =
%  \input{\projectdir QCDQED/Data/TeX/results/M_np.dat}
%  \, \text{GeV}
%  \quad 
%  [0.9389187471(83) \, \text{GeV}]_{\text{exp}}
%  \quad ,
% \end{equation}
%
\begin{align}
  M_{n}  &=
 \input{\projectdir QCDQED/Data/TeX/results/M_n.dat}
 \, \text{GeV}
 \quad 
 [0.9395654133(58) \, \text{GeV}]_{\text{exp}}
 \quad ,
 \\
  M_{p}  &=
 \input{\projectdir QCDQED/Data/TeX/results/M_p.dat}
 \, \text{GeV}
 \quad 
 [0.9382720813(58) \, \text{GeV}]_{\text{exp}}
 \quad ,
\end{align}
and
\begin{equation}
 M_n - M_p =
 \input{\projectdir QCDQED/Data/TeX/results/dM_np_feyn.dat}
 \, \text{MeV}
 \quad 
  [1.29333205(51) \text{MeV}]_{\text{exp}}
  \quad .
\end{equation}
%

% %
% We also extrapolate the isosymmetric limit $M_N$, 
% with a similar ansatz where in place of $\alpha_{EM}/{L^3}$ we put the finite volume factor for the nucleon,
% \mbox{$M_\pi^3 e^{-M_\pi L}/(M_\pi L)$}
% \cite{colangelo2010finite}.
% From that we find a value for the $\pi N$ sigma term in the chiral limit compatible with \cite{GASSER1991252, alvarez2013nucleon}:
% %
% \begin{equation}
%   \sigma_{\pi N}
%   = m_\ell \frac{\partial M_N}{m_\ell}
%   \approx M_\pi^2 \frac{\partial M_N}{M_\pi^2}
%   =
%   \input{\projectdir isoQCD/Data/TeX/results/piN_sigma_term_MeV.dat}
%   \quad .
% \end{equation}
% %

\begin{figure}
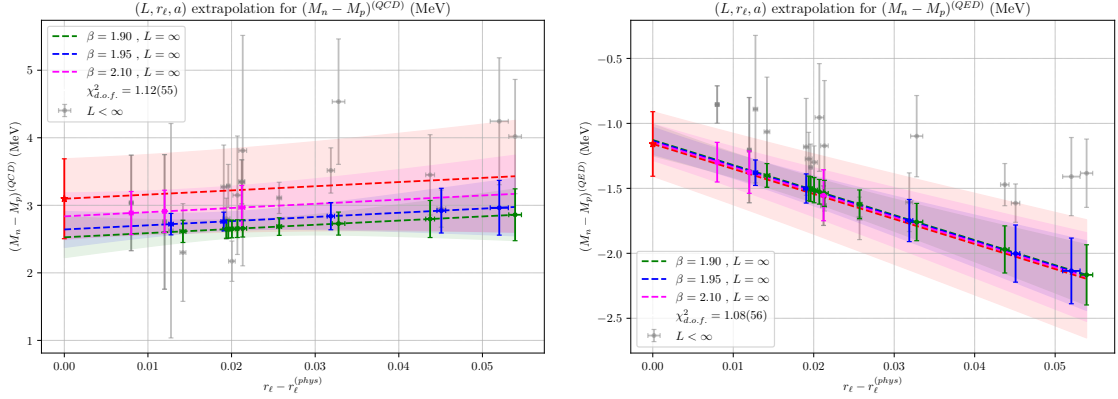

    \includegraphics[width=0.49\textwidth]{\plotsdir QCDQED/fit_Lla/dM_np_feyn_QCD.pdf}
    \includegraphics[width=0.49\textwidth]{\plotsdir QCDQED/fit_Lla/dM_np_feyn_QED.pdf}
 \caption[]{
  The same as fig. \eqref{fig:Mnp.and.dMnp.extrapolation.QCDQED} but for $(M_n-M_p)^{(QCD)}$ 
  % For the coefficients we get:
  % $A_{np}=3.05(0.61)$ MeV, 
  % $c_\ell=2.0(3.9)$.  
  % $c_a=-7.8(8.1)$, 
  % $\alpha_{EM} c_3=10.7(9.4)$, 
  and 
  $(M_n-M_p)^{(QED)}$.
  % $A_{np}=-1.03(0.24)$ MeV, 
  % $c_\ell=1.88(58)$.  
  % $c_a=-1.1(1.1)$, 
 }
 \label{fig:dMnp.QCD.QED.contributions}
 % \label{fig:dMnp.strongIB.extrapolation.QCDQED}
\end{figure}

Finally, 
we extrapolate the two contributions
$(M_n-M_p)^{(QCD)}$ 
and
$(M_n-M_p)^{(QED)}$ 
of eqs. 
\eqref{eq:dMnp.QCD.expression} 
and \eqref{eq:dMnp.QED.expression},
using the same functional form of eq. \eqref{eq:dMnp.ansatz}.
We find:
\begin{align}
  & (M_n-M_p)^{(QCD)} =  
 \input{\projectdir QCDQED/Data/TeX/results/dM_np_feyn_QCD.dat}
 \, \text{MeV}
 \quad ,
 \\
  & (M_n-M_p)^{(QED)}  = 
 \input{\projectdir QCDQED/Data/TeX/results/dM_np_feyn_QED.dat}
 \, \text{MeV}
 \quad .
\end{align}
%
% which turns out to be compatible with the saparation of \cite{borsanyi2015ab}.
In fig. \eqref{fig:dMnp.QCD.QED.contributions}
are shown the plots of the extrapolations for these two quantities.
\section{Summary and outlook}
\label{sec:summary.outlook}
In this work we've discussed the inclusion of LIBEs effects in the nucleons spectrum,
finding their masses and the difference $M_n-M_p$.
% We interpret the latter as the sum of strong IB and QED effects,
% in a prescription which turns out to be compatible with other findings in the literature \cite{borsanyi2015ab}.
% Besides we also give a prediction of nucleon sigma term $\sigma_{\pi N}$ compatible with
% \cite{GASSER1991252, alvarez2013nucleon}.

The
$O(\alpha_{EM})$ 
and 
$O(\frac{\hat{m}_d -\hat{m}_u}{\Lambda_{QCD}})$
corrections have been taken into account using the RM123 method.
%
% In our twisted mass regularization,
% this requires the introduction and tuning of counterterms.
Critical masses were fixed by the PCAC Ward Identity in order to preserve the maximal twist condition at LO. 
The physical masses are tuned using an hadronic scheme and the lattice spacing is fixed through 
$M_{\Omega^{-}}$. 

We worked in the electroquenched approximation and neglected QCD disconnected diagrams.
At our level of precision these are expected to be suppressed, giving a negligible contribution.
This assumption is confirmed by the consistency of our results with the experimental values.
For a sequent work we aim at including these diagrams,
whose effects can be known only by direct evaluation,
and provide an estimate of the various sources of systematic errors.

In the future we also aim at applying the RM123 method to neutron $\beta$ decay,
which has phase space size given by $M_n-M_p$.
A good prediction of this quantity is the preliminary step to face up to the radiative corrections in the decay width, from which one could improve the determination of radiative corrections in the CKM matrix element $V_{ud}$.
% , 
% whose best determination at the moment
% comes from $14$
% super-allowed nuclear beta decays \mbox{($0^{+} \to 0^{+}$)} 
% \cite{10.1093/ptep/ptaa104, Towner_2010}, 
% with radiative corrections given by model-dependent assumptions \cite{SIRLIN197429}.

\section{Acknowledgements}

The numerical simulations were carried out on the CINECA Tier-0 supercomputer MARCONI.
I gratefully acknowledge the CPU time provided by PRACE under the project Pra10-2693 
``QED corrections to meson decay rates in Lattice QCD'' and by CINECA under the specific initiative INFN-LQCD123.
I am also grateful to V. Lubicz, S. Simula, F. Sanfilippo, G. Martinelli and C. Tarantino for their support and fruitful discussions during the evolution of this project.

%%% bibliography

% \nocite{*} % get all citations even if not cited explicitly
\bibliographystyle{JHEP}
\bibliography{./biblio.bib}

\end{document}